\documentclass[prl,twocolumn,superscriptaddress,amsmath,amssymb]{revtex4}
\usepackage{graphicx}
\usepackage{epstopdf}
\usepackage{dcolumn}
\usepackage{bm}

\def\openone{\leavevmode\hbox{\small1\kern-3.3pt\normalsize1}}

\newcommand{\Op}[1]{\boldsymbol{\mathsf{\hat{#1}}}}

\newcommand{\Fkt}[1]{\,\mathsf {#1}}

\ifx\Tr\renewcommand{\Tr}{\Fkt{Tr}}
\else\newcommand{\Tr}{\Fkt{Tr}}
\fi

\begin{document}

\title{Generating Molecular Rovibrational Coherence by
  Two-Photon Femtosecond Photoassociation of 
  Thermally Hot Atoms}

\author{Leonid Rybak}
\affiliation{
  The Shirlee Jacobs Femtosecond Laser Research Laboratory, Schulich
  Faculty of Chemistry, 
  Technion-Israel Institute of Technology, Haifa 32000, Israel
}

\author{Saieswari Amaran}
\affiliation{
  Institute of Chemistry, Hebrew University, Jerusalem 91904, Israel
}

\author{Liat Levin}
\affiliation{
  The Shirlee Jacobs Femtosecond Laser Research Laboratory, Schulich
  Faculty of Chemistry, 
  Technion-Israel Institute of Technology, Haifa 32000, Israel
}

\author{Micha\l{} Tomza}
\affiliation{
  Department of Chemistry, University of Warsaw, Pasteura 1, 02-093 Warsaw, Poland
}

\author{Robert Moszynski}
\affiliation{
  Department of Chemistry, University of Warsaw, Pasteura 1, 02-093 Warsaw, Poland
}

\author{Ronnie Kosloff}
\affiliation{
  Institute of Chemistry, Hebrew University, Jerusalem 91904, Israel
}

\author{Christiane P. Koch}
\affiliation{Institut f\"ur Theoretische Physik,
  Freie Universit\"at Berlin,
  Arnimallee 14, 14195 Berlin, Germany}
\affiliation{Theoretische Physik,
  Universit\"at Kassel,
  Heinrich-Plett-Str. 40, 34132 Kassel, Germany}

\author{Zohar Amitay}
\email{amitayz@tx.technion.ac.il}
\affiliation{
  The Shirlee Jacobs Femtosecond Laser Research Laboratory, Schulich
  Faculty of Chemistry, 
  Technion-Israel Institute of Technology, Haifa 32000, Israel
}

\begin{abstract}
The formation of diatomic molecules with rotational and vibrational coherence
is demonstrated experimentally 
in free-to-bound two-photon femtosecond photoassociation of
hot atoms. In a thermal gas at a temperature of
1000$\,$K, pairs of magnesium atoms, colliding in their electronic
ground state, are excited into coherent superpositions of bound
rovibrational levels in an electronically excited state. 
The rovibrational coherence is probed by 
a time-delayed third photon, resulting in quantum beats in the UV
fluorescence. A comprehensive theoretical model based on \textit{ab
  initio} calculations rationalizes the generation of 
coherence by Franck-Condon filtering of collision energies and partial
waves, quantifying it in terms of an increase in quantum purity of 
the thermal ensemble. Our results open the way to coherent control of
a binary reaction. 
\end{abstract}


\maketitle

When a pair of colliding atoms is irradiated by a femtosecond pulse,
an important scenario is  photoassociation (PA) in which a chemical 
bond is "photo-catalysed"  via a free-to-bound broadband
optical transition
\cite{MarvetCPL95,BackhausCP97,VardiJCP97,BonnSci99,SalzmannPRL08,NuernbergerPNAS10}.  
Femtosecond laser pulses can be shaped in their amplitude,
phase and polarization. They are thus the basic tool for coherent
control which employs interference of matter waves to steer the
dynamics of a quantum system toward a desired outcome
\cite{RiceBook,ShapiroBook,RabitzNJP10}. 
Chemical reaction control, as a means for photo-inducing new chemical 
synthesis, is a long-standing yet unrealized goal
from the early days of coherent control \cite{ShapiroBook}.
While coherent control has been extremely successful for destructive
unimolecular processes such as ionization or dissociation
\cite{GordonARPC97,BrixnerCPC03,DantusCR04,WollenhauptARPC05,SFB450book}, control of 
associative binary reactions  still remains an unresolved puzzle. 
A controlled photoassociation step between two atoms or two
molecules has been proposed as a first step in an extended control
scheme of more complicated bimolecular reactions \cite{RonnieDancing89}.
Femtosecond photoassociation leading to chemical bond formation
is thus the prerequisite for coherent control of a binary reaction.
Femtosecond photoassociation of hot atoms (hot fs-PA) has been 
demonstrated in a single pioneering experimental study 
more than 15 years ago \cite{MarvetCPL95}, using mercury and a
single-photon UV transition. Coherent rotational molecular dynamics were
observed following the PA. 
In order to utilize hot fs-PA as a basis for coherent control
of binary reactions, it is, however, essential to
induce \textit{vibrational} coherence of the photoassociated
molecules since the vibrational degrees of freedom 
determine the fate of bond making and breaking. 
In view of a large control toolbox, it would be optimal to 
have both one-photon and multi-photon fs-PA available.
The high peak powers of femtosecond laser pulses can easily drive
multi-photon transitions, resulting in a large range of molecular
excitation energies amenable to fs-PA. Multi-photon PA allows for
accessing electronically excited molecular states that are
inaccessible in one-photon PA due to the different selection
rules. It also allows for utilizing multi-photon control strategies
that differ from one-photon control strategies due to an extended
manifold of state-to-state quantum pathways and large 
non-resonant Stark shifts \cite{YaronARPC09}.
Moreover, in view of experimental feasibility, fs-PA 
using non-resonant multi-photon transitions with near-IR or visible
photons is desirable in order to associate molecules which have their
lowest electronic transition in the deep-UV or VUV spectral range
where pulse shaping is more difficult. 
Multi-photon fs-PA thus extends the variety of molecular species and
reactions that are candidates for coherent control using shaped
pulses. 

In this Letter, we demonstrate two-photon hot fs-PA
with subsequent coherent molecular motion that is of both rotational
and vibrational nature. 
Our experimental and theoretical results are presented for 
fs-PA of hot magnesium atoms into bound magnesium dimer
molecules, i.e., Mg+Mg$\rightarrow$Mg$_{2}^{*}$,
with the thermal ensemble of Mg atoms held at a temperature of
1000$\,$K. We show  that femtosecond laser pulses 
can carve from an initial incoherent mixture a
sub-ensemble with increased quantum purity, $\Tr[\Op\rho^2]$, 
and dynamical coherences that are amenable to coherent
control.

Magnesium is experimentally well-suited for PA
studies for two main reasons. (i)
The experimental background signal originating from direct
bound-bound optical excitation of already-bound ground-state dimers 
is very small. This is due to the van der Waals nature of the 
molecular ground electronic state whose shallow well of only about 400
cm$^{-1}$ is outweighted by the rotational barrier and does not support any
bound states for most of the thermally populated partial waves.
(ii) The electronically excited states of Mg$_{2}$ with
deep potential wells are conveniently accessible
in a free-to-bound PA process via optical transitions of
two or three NIR photons. 
\begin{figure}[tb]
  \centering
  \includegraphics[width=\linewidth]{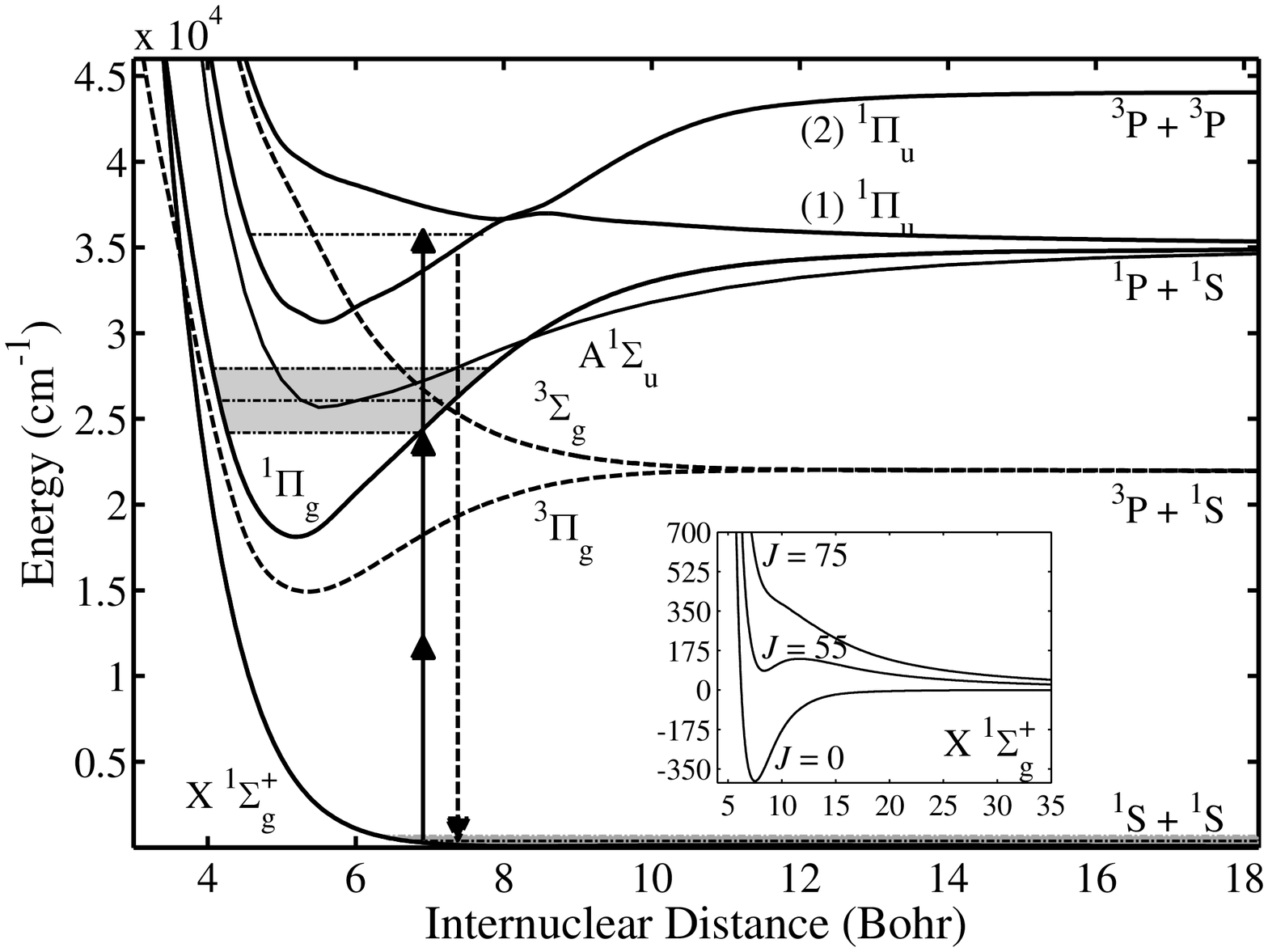}
  \caption{The excitation scheme of the two-photon femtosecond
    photoassociation of magnesium. The shaded areas indicate the
    thermally populated scattering states in the electronic ground
    state and the vibrational band of $\Pi_g$-state molecules inferred
    from the autocorrelation function. The position of the vibrational
    band \textit{above} the weak-field two-photon resonance is rationalized by
    the strong dynamical Stark shift.}
  \label{fig_pot_curves}
\end{figure}
Figure~\ref{fig_pot_curves} displays the relevant Mg$_{2}$ potential energy curves
and the present excitation scheme. 
The inset of the figure 
shows the ground  $X^{1}\Sigma_{g}^{+}$ state for three different values of 
partial waves, $J$, of the colliding Mg atoms:
Already for $J \approx 50$, the $X$-state does not
support any bound levels. The temperature of $T=1000\,$K corresponds to a
collision energy $k_BT/2$ of 
$\sim 350\,$cm$^{-1}$. Pairs of Mg atoms
colliding in the $X^{1}\Sigma_{g}^{+}$ electronic ground state  of
Mg$_{2}$ are excited via a broadband free-to-bound non-resonant two-photon
transition into coherent superpositions of bound
rovibrational levels in the 
electronically excited $(1)^{1}\Pi_{g}$ state.
The non-resonant two-photon coupling is provided by 
off-resonant dipole transitions to intermediate states of
\textit{ungerade} symmetry with the largest contribution coming from 
the lowest states, $A^{1}\Sigma_{u}^{+}$, $(1)^{1}\Pi_{u}$ and
$(2)^{1}\Pi_{u}$. PA is induced
by pulses of 70$\,$fs transform-limited duration,
840$\,$nm central wavelength, linear polarization, and a
transform-limited peak intensity of about
5$\times$10$^{12}$~W/cm$^{2}$. 
A time-delayed probe pulse, identical to the pump pulse,
probes the dynamics of the excited dimers via a one-photon excitation to
the higher lying $(1)^{1}\Pi_{u}$ electronic state.
This state has a strong one-photon allowed transition to the
$X^{1}\Sigma_{g}^{+}$ ground state with a lifetime of a few ns. The
intensity of the resulting UV fluorescence  at 285-292$\,$nm
is measured as a function of the pump-probe time delay
$\tau$, yielding the pump-probe signal transient. 
The experiments are conducted in a static cell with argon buffer gas
and a pressure of the Mg vapor of $\sim 5\,$Torr. The sample is
irradiated at a repetition rate of 1$\,$kHz by the 
pump and probe laser beams. 
UV fluorescence emitted toward the laser-beam entrance to the cell is
collected at a small angle from the laser-beam axis using an
appropriate optical setup. 

\begin{figure}[tb]
  \centering
  \includegraphics[width=\linewidth]{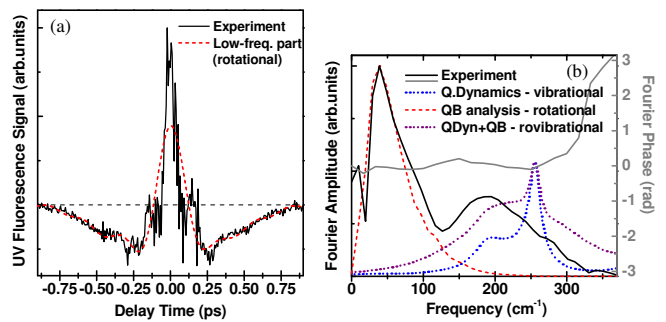}
  \caption{(color online) Two-photon femtosecond
    photoassociation of hot magnesium atoms:
    (a) Experimental pump-probe signal showing the measured UV
    fluorescence intensity as a function of the pump-probe delay
    (black line). 
    The red line corresponds to rotational dynamics only.
    It results from a low-pass step filter set to include only 
    frequencies smaller than 4$\,$THz or $133.42\,$cm$^{-1}$.
    (b) Fourier spectrum of the experimental pump-probe trace (solid
    black and grey lines). 
    Superimposed are the theoretical amplitude spectra corresponding to 
    purely vibrational dynamics (blue dotted line), purely rotational
    dynamics (dashed red line) and vibrational dynamics including
    rotational broadening (purple dotted line).
    }
  \label{fig_exp_results}
\end{figure}
Figure~\ref{fig_exp_results} presents experimental results for 
two-photon fs-PA of hot magnesium atoms, showing 
the measured pump-probe transient in Fig.~\ref{fig_exp_results}(a) and
its Fourier Transform (FT) spectrum in Fig.~\ref{fig_exp_results}(b).
The time-dependent pump-probe transient signal, symmetric around
$\tau=0$ since pump and probe pulses are identical, is 
observed also for delay times 
where the photoassociating pump pulse has already ended.
As corroborated by the calculations presented below, observation
of this time-dependence demonstrates that 
(i) molecules have been formed 
in a non-stationary coherent state, 
i.e., in a coherent superposition of molecular rovibrational states,
and
(ii) the generated ensemble of photoassociated molecules must be of
increased purity as compared to the initial thermal, incoherent atomic
ensemble -- without an increase in purity, 
no coherence could be observed.
The FT spectrum of the  experimental pump-probe signal,
cf. Fig.~\ref{fig_exp_results}(b),  
is composed of a narrow low-frequency peak at around 38.9~cm$^{-1}$
and a broad high-frequency peak located at 189.6~cm$^{-1}$.
It shows a zero spectral phase, confirming that pump and
probe excitation occur at the same $R$, cf. Fig.~\ref{fig_pot_curves}.  
Below, we show the low-frequency peak to be associated to coherent
rotational dynamics and the high-frequency peak to coherent
vibrational dynamics. 

To interpret the experimental data, we have constructed a
comprehensive theoretical model. 
State-of-the-art \textit{ab initio} methods have been employed to
calculate the Mg$_{2}$ potential energy curves presented in
Fig.~\ref{fig_pot_curves} as well as
the corresponding one-photon and two-photon transition dipole moments,
dynamical Stark shifts, and spin-orbit interaction, non-adiabatic and
radial couplings in the electronically excited states.
The short-range part of the potentials and transition
moments were calculated within the framework of the 
equation of motion (reponse) coupled cluster
method restricted to single and double
excitations~\cite{Sekino,Jorgensen:90} in a large basis set. 
The long-range part of the potentials as well as the
non-adiabatic, 
radial and spin-orbit couplings have been obtained using
multi-reference configuration interaction. A high accuracy of
the computed potential energy curves is confirmed by a good
agreement of the calculated spectroscopic constants with their
experimental values.
Our \textit{ab initio} data are by far much more accurate than
previous results reported in 
Refs.~\cite{Czuchaj_TheoChemAcc_107_27_2001}. 
The PA dynamics are simulated non-perturbatively
by solving the
time-dependent Schr\"odinger equation on a multi-surface grid
including all non-adiabatic couplings and one-photon and two-photon
transition dipole operators.
It is crucial to correctly represent the initial thermal
ensemble of scattering atoms
(as opposed to a single state which is always coherent).  
The incoherent initial state is described by 
the density operator in the
canonical ensemble, $\Op\rho(t=0)=\exp[-\beta \Op H_g]/Z$ with
$\beta=1/k_BT$, $\Op H_g$ the Hamiltonian for nuclear motion in the
ground electronic state, and $Z$ the partition function, 
$Z=\Tr[\exp[-\beta \Op H_g]]$
It is constructed using random-phase
wavefunctions \cite{Gelmanandronnie}.  
Details of the electronic structure
calculations and the random-phase thermal wavefunction approach are
reported elsewhere~\cite{theory}.

Separating rotational and vibrational motion, 
the PA dynamics is simulated by propagating many
realizations of random-phase thermal wavefunctions 
under the pump and probe pulses
for each value, $J$, of the incoming partial waves. 
Computing the expectation value of an observable for each
propagated random-phase wavefunction and incoherently summing all 
individual expectation values yields the 
thermally averaged time-dependent expectation value.
For example, the calculated spectrum shown in
Fig.~\ref{fig_exp_results}(b) (blue dotted line) 
is obtained by Fourier transforming  
the thermally averaged time-dependent population of the $^{1}\Pi_{u}$
states. 
Since the full pump-probe simulation for many random-phase
realizations and all thermally populated partial waves $J$
is numerically rather demanding, 
our calculations account only for optical transitions of
$\Delta J=0$. 
Our quantum dynamical results therefore reflect the
dependence of hot fs-PA  on the partial waves, $J$,
but the calculated dynamics is of purely \textit{vibrational} nature.

\begin{figure} [tb]
  \includegraphics[width=\linewidth]{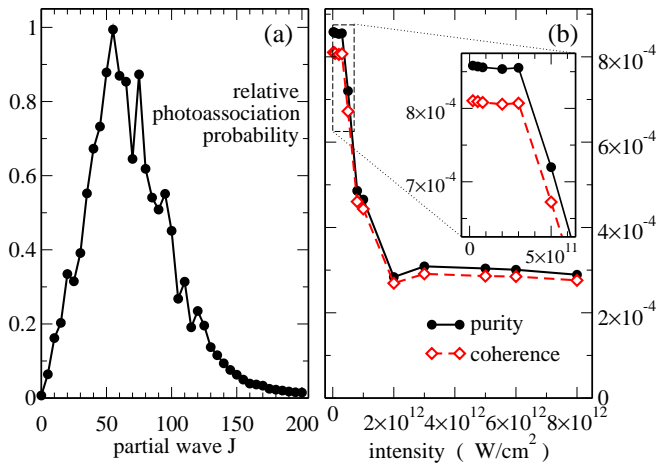}
  \caption{\label{fig_theoretical_results}
    (a) Calculated relative PA probability as a function of incoming
    partial wave $J$.
    (b) Quantum purity, $\Tr[\Op \rho(t_{final})^2]$ and dynamical
    coherence of the sub-ensemble of molecules in the $(1)^1\Pi_g$
    state vs laser intensity} 
\end{figure}

Comparing the experimental spectrum to the calculated one (black solid and blue
dashed lines in Fig.~\ref{fig_exp_results}(b)), we find the 
high-frequency peak to be due to coherent vibrational dynamics. The
range of populated vibrational levels in the $^1\Pi_g$ state is indicated by
the grey shaded area in Fig.~\ref{fig_pot_curves}. 
We approximately account for the rotational dynamics by 
carrying out a quantum-beat analysis:
We assume rovibrational $^1\Pi_g$ levels 
with $J'=J, J\pm 1,J\pm 2,J\pm 3, J\pm 4$ to get excited by the pump
pulse from each incoming partial wave $J$.  
Each pair of rovibrational $^1\Pi_g$ levels, $(v'_1,J'_1)$,
$(v'_2,J'_2)$,
with  $v'_2=v'_1,v'_1+1$ and $|J'_2-J'_1|=2,4$
contributes to the FT spectrum an amplitude $P_{^1\Pi_g}(J)$ 
at the frequency $\omega_{QB}=(E_{v'_1,J'_1}-E_{v'_2,J'_2})/\hbar$. 
$P_{^1\Pi_g}(J)$ is the $^1\Pi_g$ state population obtained in the
vibrational dynamics simulation for each incoming $J$, shown in
Fig.~\ref{fig_theoretical_results}(a). The result of the quantum-beat
analysis is  
shown by the red dashed line in Fig.~\ref{fig_exp_results}(b) and 
describes the low-frequency peak in the experimental spectrum very
well. We thus find this peak to fit pairs of $^1\Pi_g$ levels with
$v'_1=v'_2$ and $J'_1,J'_2 \approx 40\dots 100$. In particular, the
quantum beat frequency corresponding to the maximum of the peak
matches only pairs with $J'_{1,2} \approx 50 \ldots 60$.
Combining our approximate treatment of the rotational dynamics with
the full quantum simulation of the vibrational dynamics leads
to rotational broadening of the vibrational peak, cf. purple dotted
line in Fig.~\ref{fig_exp_results}(b). We attribute the remaining
small discrepancy between the experimental and the theoretical 
spectrum to the separation of rotational and vibrational dynamics 
and the uncertainty of the \textit{ab initio} calculations. In
particular, the sharp feature in the calculated vibrational spectrum is
due to the non-adiabatic couplings between the $^1\Pi_u$ states which
carry the largest error bar of the otherwise very accurate \textit{ab
  initio} results.
To summarize, the peaks observed in the experimental spectrum originate
from coherent excitation and probing of pairs of rovibrational 
levels $(v'_1,J'_1)$ and $(v'_2,J'_2)$ with high rotational quantum
numbers in the $(1)^{1}\Pi_{g}$
electronically excited state. 

As illustrated in the inset of Fig.~\ref{fig_pot_curves},
for values of $J \gtrsim 50$, the $X^{1}\Sigma_{g}^{+}$ ground state of Mg$_{2}$
does not support any bound molecular level.
We therefore deduce from our quantum beat analysis 
that indeed the majority of the bound excited Mg$_{2}$ molecules, that 
we probe experimentally, have been generated 
in a free-to-bound coherent PA process.
This conclusion of free-to-bound PA of high-$J$ molecules
is further supported by the calculated relative PA probabilities,
shown in Figure~\ref{fig_theoretical_results}(a), as a function of the
incoming partial wave, $J$. Maximum PA probability is observed 
for $J=55$, for which the electronic ground state does not support
bound levels, cf. Fig.~\ref{fig_pot_curves}.
A careful analysis shows that about 80 per cent of the total signal is
due to free-to-bound PA~\cite{theory}. 

In order to rationalize and quantify the generation of the coherence, that is
observed in Fig.~\ref{fig_exp_results}, out of a completely incoherent
initial ensemble, we consider the change in quantum purity,
$\Tr[\Op\rho^2]$, due to the femtosecond laser pulses. 
The initial purity, determined by the temperature,
$T=1000\,$K, and density,  $\rho=4.8\cdot10^{16}\,$atoms/cm$^3$, of the
experiment, 
is estimated in terms of the purity of a single atom pair represented
in our computation and the probability, $p_2$, of finding two atoms in our
computation volume, $\mathcal{V}$. 
Given $\rho$ and $\mathcal{V}=4/3\pi r^3=6.33 \cdot 10^{-21}\,$cm$^{3}$,
we find $p_2=9.2 \cdot 10^{-8}$. 
The purity of a single atom pair in our
computation box is given by the vibrational purity weighted by the 
probability, $P_J$, for occupation of the sub-ensemble with 
angular momentum $J$.
For an atom pair in their electronic ground state, 
the purity becomes 
${\cal P}^g = p_2^2 \sum_J P_J^2 {\cal P}^g_J$ where  
${\cal P}^g_J= \Tr [(e^{-\beta \Op H_g^J})^2]/ Z_J^2$, and $\Op H_g^J$
and $Z_J$ are the Hamiltonian for vibrational motion and partition
function, respectively,  of partial wave $J$. Evaluating 
${\cal P}^g_J$ with thermal random-phase wavefunctions, 
we obtain ${\cal P}^g=2.6 \cdot 10^{-20}$ for the initial purity.  
For the ensemble of molecules in the electronically excited $^1\Pi_g$ state, 
the density operator is given by $\Op\rho^e=\sum_J P^e_J \Op\rho^e_J$.
Here, $P^e_J$ is the probability for occupation of the excited state
sub-ensemble with angular momentum, $J$,
cf. Fig.~\ref{fig_theoretical_results}(a). 
$\Op\rho^e_J$ is the normalized density operator of the excited state
$J$ sub-ensemble which is constructed by incoherently averaging 
dyadic products of the propagated thermal wavefunctions,
$|\psi_{n,J}(t_{final})\rangle\langle\psi_{n,J}(t_{final})|$, 
over all random phases $n$ and partial waves $J$ ($t_{final}$ denotes
the time when the pump pulse is over). 
The purity for the excited state sub-ensemble is then given by 
${\cal P}^e =\sum_J (P^e_J)^2 \Tr [(\Op\rho^e_J)^2]$.
We obtain a purity ${\cal P}^e=3.04 \cdot 10^{-4}$ for the molecular
sub-ensemble in the $^1\Pi_g$
excited state for the experimental pulse parameters. We thus observe a
dramatic increase in the quantum purity, $\Tr[\Op\rho^2]$, 
induced by the femtosecond laser pulse. 
The underlying physical mechanism can be viewed as "Franck-Condon
filtering": 
for a given initial $J$ value there is only a limited range of
collision energies that allow the colliding pair to reach
the Franck-Condon window for PA located at short internuclear
distances \cite{BackhausCP97}. 

In order to obtain a quantitative estimate of the degree of
distillation achieved by the fs-PA process, 
we have calculated the purity of ensemble of photoassociated
molecules in the $^{1}\Pi_{g}$ state for a range of laser intensities,
cf. Fig.~\ref{fig_theoretical_results}(b). For weak fields, the purity
is roughly constant as a function of intensity and almost three times
larger than the purity obtained for the intensity of $5\times
10^{12}\,$W/cm$^2$ used in the experiment. As intensity is increased,
a sudden drop in the purity is observed which levels off at large
intensities. We attribute this drop to power broadening for strong
fields, which brings more atom pairs into the Franck-Condon window for
PA. The purity increase with respect to the initial ensemble is
nevertheless many orders of magnitude for both weak and strong field. 
To further analyze the generation of quantum coherence, 
it is possible to separate static and dynamic
contributions, $\Op\rho = \Op\rho_{stat} +
\Op\rho_{dyn}$. This is most easily
achieved in the energy representation where the static (dynamic) part
corresponds to the diagonal (off-diagonal) matrix elements.
The dynamical contributions are quantified by the
coherence measure ${\cal C} =
\Tr[\Op\rho^2_{dyn}(t)] < \Tr[\Op\rho^2(t)]$~\cite{BaninJCP94}. 
Tracing the coherence measure of the excited state, ${\cal C}^e$,
as a function of laser intensity (red open diamonds in
Fig.~\ref{fig_theoretical_results}(b)) 
shows that most of
the purity comes from the dynamical contribution (${\cal C}^e =
2.86 \cdot 10^{-4}$ for 
$I=5\times 10^{12}\,$W/cm$^2$).

In summary, we have demonstrated femtosecond photoassociation of hot
atoms using non-resonant two-photon transitions driven by NIR light. 
Quantum beats in the observed UV fluorescence reflect both vibrational
and rotational coherence of the photoassociated molecules. 
Our experiment is the first to demonstrate generation of 
vibrational coherence in a light-induced binary reaction. 
This is rationalized by our theoretical model in terms of 
Franck-Condon filtering, yielding an increased quantum purity of the
sub-ensemble of the reaction products, the photoassociated molecules, 
as compared to that of the reaction partners, the atoms. 
Vibrational coherence is a prerequisite for further control of the
reaction products. Our work thus opens the way toward coherent control
of a binary reaction. 

\begin{acknowledgments}
This research was supported by The Israel Science Foundation (grant
No.~1450/10), The James Franck Program in Laser Matter
Interaction, the Polish Ministry of Science and
Higher Education (project N N204 215539), the Foundation for Polish
Science MPD Programme co-financed by 
the EU European Regional Development Fund, 
and the Deutsche Forschungsgemeinschaft.
\end{acknowledgments}


\end{document}